\documentclass[aps,pre,superscriptaddress,twocolumn,longbibliography]{revtex4-2}

\usepackage{graphicx}
\usepackage{dcolumn}
\usepackage{bm}
\usepackage{epstopdf}
\usepackage{algorithm}
\usepackage{algpseudocode}
\usepackage[colorlinks=true,linkcolor=black,anchorcolor=black,citecolor=black,filecolor=black,menucolor=black,runcolor=black,urlcolor=black]{hyperref}
\usepackage{amsmath,amsthm,amssymb}

\usepackage{float}
\usepackage{epsfig}
\usepackage{bm}
\usepackage{mathrsfs}
\usepackage{multirow}
\usepackage[all]{xy}
\usepackage{pbox}
\usepackage{verbatim}
\usepackage{braket}
\usepackage{mathtools}
\usepackage{bm}
\usepackage{tikz}
\usepackage{xcolor}
\usepackage{xfrac}

\newcommand{\er}[1]{Eq.\eqref{#1}}
\newcommand{\fr}[1]{Fig.\ref{#1}}
\newcommand{\algr}[1]{Alg.\ref{#1}}

\newcommand{\Er}[1]{Equation~\eqref{#1}}

\newcommand{\beq}{\begin{equation}}
\newcommand{\eeq}{\end{equation}}

\newcommand{\params}{\bm{\theta}} 
\newcommand{\trajparams}{\bm{\Theta}} 

\DeclareSymbolFont{bbold}{U}{bbold}{m}{n}
\DeclareSymbolFontAlphabet{\mathbbold}{bbold}
\newcommand{\idty}{\mathbbold{1}}

\begin{document}

\title{Training neural network ensembles via trajectory sampling}
\author{Jamie F. Mair}
\email{jamie.mair@nottingham.ac.uk}
\affiliation{School of Physics and Astronomy, University of Nottingham, Nottingham, NG7 2RD, UK}
\author{Dominic C. Rose}
\affiliation{Department of Physics and Astronomy, University College London, Gower Street, London WC1E 6BT, UK}
\author{Juan P. Garrahan}
\affiliation{School of Physics and Astronomy, University of Nottingham, Nottingham, NG7 2RD, UK}
\affiliation{Centre for the Mathematics and Theoretical Physics of Quantum Non-Equilibrium Systems,
University of Nottingham, Nottingham, NG7 2RD, UK}

\date{\today}

\begin{abstract}
In machine learning, there is renewed interest in neural network ensembles (NNEs), whereby predictions are obtained as an aggregate from a diverse set of smaller models, rather than from a single larger model. Here, we show how to define and train a NNE using techniques from the study of rare trajectories in stochastic systems. We define an NNE in terms of the trajectory of the model parameters under a simple, and discrete in time, diffusive dynamics, and train the NNE by biasing these trajectories towards a small {\em time-integrated} loss, as controlled by appropriate counting fields which act as hyperparameters. We demonstrate the viability of this technique on a range of simple supervised learning tasks. We discuss potential advantages of our trajectory sampling approach compared with more conventional gradient based methods.
\end{abstract}

\maketitle

\section{Introduction}

The traditional approach in machine learning (ML), once the architecture of a model is defined (say a neural network, or NN, composed of layers of coupled neurons), is to learn one set of parameters (say the couplings and biases between the neurons) as optimally as possible from the data; for reviews see e.g.~\cite{Goodfellow2016,Mehta2019}. This is done by minimising a (suitably regularised) objective or loss function of the parameters over a training data set \cite{Goodfellow2016,Mehta2019}. This optimisation amounts to gradient descent in the landscape defined by the loss function and the training data, often supplemented with tricks that help to speed up convergence, such as adding inertia or stochasticity to the optimisation dynamics \cite{Bottou2003,Kingma2014}. The properties of the training dynamics are controlled by so-called hyperparameters \cite{Goodfellow2016,Mehta2019}. In this approach, at the end of the learning process one gets a single trained model that is then is used to make inferences \cite{Goodfellow2016,Mehta2019}.  

However, there has been recent interest \cite{Wen2020,Wang2020} in an alternative approach, where rather than a single model, one trains a set of models. Such ensemble or committee of models \cite{Hansen1990,Perrone1992,Krogh1994,Opitz1999,Lakshminarayanan2017,Huang2017,Kurutach2018} offers several advantages. Since training by gradient descent can converge to different solutions in a complex loss landscape, 
training an ensemble of models starting from different initial seeds gives a set of equivalent yet distinct models. Obtaining inferences as the mean or as a majority consensus of the ensemble members can therefore provide better estimates and attenuate uncertainty. Furthermore, an ensemble of smaller models may be more computationally efficient than a single large model, both to train and to run \cite{Wang2020}.

Here, we introduce a method to define and train neural network ensembles (NNEs) that is based on ideas from the sampling of rare stochastic trajectories. 
From a statistical mechanics perspective, training one ML model with stochastic gradient descent is equivalent to thermal annealing \cite{Whitelam2021}, where the dynamical variables are the model parameters, the training data plays the role of quenched disorder, the loss is the energy, and stochasticity comes from a thermal bath: at low temperature, an annealing dynamics of the ML parameters, such as Monte-Carlo, will converge to a state of low energy and therefore low loss. 
Analogously, we can think of a NNE in terms of a {\em trajectory of models}: 
given some dynamics of the parameters, the set of configurations visited over a period of time defines the ensemble. If we require the trajectory to have low {\em time-integrated loss} then we can obtain a well-trained NNE. This procedure can be implemented with modern trajectory sampling techniques
\cite{Bolhuis2002,Hedges2009,Garrahan2018,Jack2019}, and as we show below, it
is a viable way to define and train ML ensembles. 

The paper is organised as follows. In Sec.~II we review basic concepts about neural networks and NN ensembles. 
In Sec.~III we introduce our method 
of defining NNEs in terms of trajectories of a stochastic dynamics. Section IV provides exact results for the simple case where the NN architecture is that of a linear perceptron. In Sec.~V we show how to train NNEs by means of transition path sampling. In Sec.~VI we illustrate our method with application to the linear perceptron, to a two-dimensional loss landscape, and to the textbook problem of classifying handwritten digits in the MNIST data set. Section VII gives our conclusions and outlook.

\section{Neural Networks and Neural Network Ensembles}

\subsection{Neural Networks}

Neural networks are computational models which are commonplace throughout ML~\cite{Goodfellow2016,Mehta2019}. They are used as function approximations on problems where the exact structure of the mapping between input and output is unknown. A standard NN maps an input data vector $\bm{x}$ to an output vector $\bm{y}$, via $\bm{y}=f(\bm{x})$. The structure of a standard feed-forward NN can be expressed as follows:
\begin{equation}
    \bm{h}^{(i)}=\sigma^{i}(\bm{W}^{(i)}\bm{h}^{(i-1)}+\bm{b}^{(i)}),
\end{equation}
where
\begin{align}
    \bm{h}^{(0)}\ = \bm{x} \\
    \bm{h}^{(N)}\ = \bm{y}.
\end{align}
The model is parameterised by the weight matrices and the bias vectors, denoted by $\bm{W}^{(i)}$ and $\bm{b}^{(i)}$ respectively. The number of layers in the model is denoted by $N$, not counting the input layer. The activation function $\sigma^{i}$ is any non-linear function, such as a hyperbolic tangent or a rectified linear unit (ReLU).
The exception is that the output layer can be allowed to have a linear activation function, to not limit the range of outputs a model, depending on the type of function being approximated.
We denote the parameters of the model $\params$ and the function mapping input to output in the model as $f_{\params}$. We treat $\params$ as a flat vector containing all the parameters of the weight matrices and bias vectors. 

One can define an ML problem via the specification of a loss function, $L(\params)$, which we view as a function of the model parameters. This translates the problem of finding the optimal parameters that define the model into one of minimizing the loss. Most modern ML problems use a NN as the function approximation, since the structure allows for efficient calculation of the gradient of the loss, with respect to the parameters of the model. These gradients can then be used to reduce the loss through a range of techniques for updating the parameters\cite{Kingma2014, Riedmiller1993,Sutskever2013,Graves2013}; all of which are variants of basic gradient descent, where parameters are updated via
\begin{equation}
    \params \leftarrow \params - \alpha \nabla_{\params} L(\params),
\end{equation}
where $\alpha$ is a learning rate which is used to tune the size of each update.

An alternative to gradient based optimization is gradient-free optimization, such as neuroevolution \cite{Whitelam2021,Morse2019,Rodrigues2020}.
Such approaches are usually based on randomised Monte-Carlo changes to the model, which may modify both the parameters and the model structure.
Each modification is evaluated using the loss function, and probabilistically accepted or rejected according to some chosen criteria. In particular, thermal annealing with Monte-Carlo can be shown \cite{Whitelam2021} to be analogous to gradient descent.

\subsection{Neural Network Ensembles}

Instead of a single trained NN, one can use a collection of trained NNs to make an inference. The different models in such a NN ensemble \cite{Wen2020,Wang2020,Hansen1990,Perrone1992,Krogh1994,Opitz1999,Lakshminarayanan2017,Huang2017,Kurutach2018} need not share a common structure, but it is important to define a way of combining the individual predictions from the constituent models to form an aggregate prediction. For example, in the case of classification problems, one can give each model in the ensemble a ``vote'' for each prediction, and the final prediction is the most voted for option. Alternatively, when the inference is a score, the aggregate score could be the mean over the ensemble. 

We will focus on the more standard case of NNEs of models with identical structure but differing parameters. 
We denote the ensemble by the set of parameters in each of the models, $\trajparams = \{\params_{t}\}_{t=1}^\tau$, where $\tau$ is the number of models in the ensemble.
A loss for the entire ensemble is defined as $\mathcal{L}(\trajparams) = \sum_t L(\params_{t})$, where $L(\params_{t})$ is the loss of a single model with parameters $\params_{t}$.

\section{Neural Network ensembles as trajectories}

Our approach is to construct a probability distribution which puts a high weight on ensembles of models with low total loss, from which we then sample effective ensembles. This is achieved by taking inspiration from the trajectory ensemble methods for many-body stochastic processes, based on large deviations \cite{Lecomte2007,Garrahan2007,Garrahan2009,Touchette2009,Garrahan2018,Jack2019}. In what follows we show how to construct NNEs in terms of appropriately sampled stochastic trajectories of the parameters that define the individual NNs of the ensemble.

We can think of our trajectory NNE construction in two alternative but equivalent ways. The first one is as follows. We think of a NNE as a time ordered {\em trajectory} of models, $\trajparams = \theta_1 \to \theta_2 \cdots \to \theta_\tau$, where the sequence of models is generated by a stochastic dynamics of the model parameters. We start with an {\em unbiased} dynamics, which for simplicity we choose as a discrete in time random walk in the space of parameters, 
with symmetric Gaussian steps of variance $\sigma^2$. That is, under this unbiased dynamics, the trajectory $\trajparams$ has probability
\begin{align}
    \label{unbiased1}
    P_{\sigma}(\trajparams) 
        = 
        \frac{1}
        {\mathcal{Z}_{\tau}(\sigma)}
        p(\theta_1)
        \prod_{t=1}^{\tau-1} 
            \exp \left [ -\frac{1}{2 \sigma^2}(       \params_{t}-\params_{t+1})^2 \right ],
\end{align}
Here $p(\theta_1)$ is the initial probability of the parameters of the first model, and $\mathcal{Z}_{\tau}(\sigma)$ a normalisation. A NNE generated in this way would have an arbitrary loss. In order to generate a useful NNE we wish to select trajectories with low {\em time-integrated loss}
\begin{align}
    \mathcal{L}(\trajparams) 
        = 
        \sum_{t=1}^\tau
            L(\params_t)
    \label{tiloss}
\end{align}
This is done by {\em tilting} \er{unbiased1} \cite{Lecomte2007,Garrahan2007,Garrahan2009,Touchette2009,Garrahan2018,Jack2019}
\begin{align}
    \label{biased1}
    P_{\sigma,s}(\trajparams) 
        = &
        \frac{1}
        {\mathcal{Z}_{\tau}(\sigma,s)}
        p(\theta_1) e^{-s L(\theta_1)}
        \nonumber \\
        &
        \times 
        \prod_{t=1}^{\tau-1} 
            \exp \left [ -\frac{1}{2 \sigma^2}(       \params_{t}-\params_{t+1})^2 \right ]
            e^{-s L(\theta_t)}
        \\
        = &
        \frac{p(\theta_1)}
        {\mathcal{Z}_{\tau}(\sigma,s)}
        e^{-s {\mathcal L}(\trajparams)}
        \prod_{t=1}^{\tau-1} 
            \exp \left [ -\frac{1}{2 \sigma^2}(       \params_{t}-\params_{t+1})^2 \right ],
        \nonumber
\end{align}
where the normalising factor is the dynamical partition sum, given by
\begin{align}
    \label{Z}
    \mathcal{Z}_{\tau}(\sigma,s) 
        =
        \int d\trajparams
        \; 
        &
        p(\theta_1)
        e^{-s {\mathcal L}(\trajparams)}
        \nonumber \\
        & 
        \times
        \prod_{t=1}^{\tau-1} 
            \exp \left [ -\frac{1}{2 \sigma^2}(       \params_{t}-\params_{t+1})^2 \right ].
\end{align}
In $P_{\sigma,s}(\trajparams)$ the unbiased probabilities of \er{unbiased1} are re-weighted by an exponential factor in the time-integrated loss. 

The probability given by \eqref{biased1} for a trajectory, or NNE, is controlled by the ``hyperparameters'' $\sigma$ and $s$. The first one determines how different subsequent models in the trajectory are, since larger $\sigma$ corresponds to larger steps in the unbiased diffusive dynamics in parameter space (that is, $\sigma$ is the conjugate to the ``dynamical activity'' of the trajectory \cite{Garrahan2007,Maes2020}). The second hyperparameter controls (i.e., is conjugate to) the time-integrated loss, since the larger $s$, the lower the total loss in the NNE.

Our aim is to sample NNEs from \er{biased1} at large enough $s$ and, therefore, low enough overall ensemble loss. While generating trajectories with the unbiased probability \er{unbiased1} is done straightforwardly by simply running a diffusive dynamics on the parameters, obtaining trajectories compatible with \eqref{biased1} is more difficult. Difficulty arises as the tilted trajectories correspond to an atypical subset of trajectories of those generated by the diffusive dynamics, one which is exponentially suppressed in $\tau$ and in the number of parameters with respect to the typical trajectories. Nevertheless, as we show below, such subset can be efficiently accessed by means of rare event sampling techniques.

\subsection{Connection to stochastic gradient descent}

A second way to see the re-weighted, or biased, trajectories in \er{biased1} connects to more traditional approaches for NN training related to stochastic gradient descent. In the limit of $\tau=1$ the NNE is simply a single model with probability, from \er{biased1}, 
\begin{align}
    \label{eq:single-neuroevolution-sample}
    p_s(\params) =
    \frac{1}{\mathcal{Z}_s}
    e^{-s L(\params)} ,
\end{align}
where in the following we consider only $s>0$, which allows us to normalise this probability distribution. \Er{eq:single-neuroevolution-sample} is the equilibrium probability for a stochastic dynamics obeying detailed balance with respect to energy $L(\params)$ at 
inverse temperature $s$, and where $\mathcal{Z}_s = \int d{\params} e^{-s L(\params)}$. One such dynamics is so-called neuroevolution \cite{Whitelam2021}, in which the parameters of the model are updated by proposing random Gaussian increments, and accepting them with a Metropolis criterion $\min(1, e^{-s \Delta L})$, where $\Delta L$ is the change in loss. This process can be shown to be equivalent to stochastic gradient descent when averaged over many runs \cite{Whitelam2021}. 
For the case of many models, $\tau >1$, and $\sigma \to \infty$, \er{biased1} describes $\tau$ uncoupled models equilibrated under neuroevolution (or similar thermal annealing of the individual losses at inverse temperature $s$), 
\begin{align}
    \label{NEtau}
    P_{\infty,s}(\trajparams) 
        = &
        \frac{1}
        {\mathcal{Z}_\tau(\infty,s)}
        e^{-s {\mathcal L}(\trajparams)}
\end{align}
with $\mathcal{Z}_\tau(\infty,s) = (\mathcal{Z}_s)^\tau$. While \er{NEtau} does describe an NNE, all the models in the ensemble are distributed identically and independently, cf.\ \er{tiloss}, and the expected loss per model in the NNE is the same as the expected loss of an individual model under 
\er{eq:single-neuroevolution-sample}. 

In order to reduce the NNE loss, one has to couple the different models in \er{NEtau}. 
This is precisely what \er{biased1} does when $\sigma < \infty$: in this case $\sigma$ controls how much each successive model in the ensemble is allowed to differ from the previous one. That this will reduce the total loss of the ensemble can be seen from the fact that in the limit of vanishing $\sigma$, all the models have to be the same and \er{biased1} becomes
\begin{align}
    \label{NEtau_vanishing_sigma}
    P_{0,s}(\trajparams) 
        \propto &
        \exp[-s \tau L(\params_1)]
        \prod_{t=2}^\tau
        \delta(\params_t-\params_1)
\end{align}
so that the NNE is equivalent to a single model equilibrated as in \er{eq:single-neuroevolution-sample}
but at a lower temperature $(s \tau)^{-1}$, and thus a much lower average loss.

\subsection{Training strategy}

From studies of other problems with complex optimisation landscapes, such as glasses and spin glasses, it is well known that directly attempting to access low temperature states is riddled with slow convergence problems. This makes training an NNE by simply reducing the temperature, cf.\ \er{NEtau}, impractical. In contrast, the combination of the hyperparameters $s$, conjugate to the loss, and $\sigma$, conjugate to the dynamical activity can help overcome the convergence problem, as shown in large deviation studies of glassy systems \cite{Hedges2009}. 

Sampling trajectories distributed according to the tilted measure \er{biased1} is our method of training 
NNEs. The technical problem is that while trajectories are easy to generate via the diffusive dynamics that defines \er{unbiased1}, they are notoriously difficult to generate for \er{biased1}, as this represents a subset of rare diffusive trajectories with atypical time-integrated loss for $s>0$. To do this, we will employ transition path sampling (TPS) \cite{Bolhuis2002}, supplemented by convergence-enhancing tricks for trajectory proposals, in order to efficiently sample trajectories from \er{biased1}, as explained in detail in Sec.~V. 

\section{Exact results for a linear perceptron}
\label{sec:exact-linear-perceptron}

As an elementary and analytically tractable example, we consider a regression problem using a linear perceptron with a mean-squared error (MSE) loss function. 
The linear perceptron is a model given by $\bm{y}=W\bm{x}+\bm{b}$, which we simplify by defining a modified input vector $\bm{\tilde{x}}^T=[\bm{x}^T \ 1]$, in turn defining an expanded weight matrix $\tilde{W}=[W\ \bm{b}]$. 
This simplifies the model to $\bm{y}=\tilde{W}\bm{\tilde{x}}$, then $\bm{y} \in \mathbb{R}^{k}$, $\bm{\tilde{x}} \in \mathbb{R}^{(d+1)}$ and $\tilde{W} \in \mathbb{R}^{k \times (d+1)}$, where $k$ is the number of target output dimensions and $d$ is the number of dimensions of the feature vector. 
Given $N$ target labels, $\bm{y'}$, corresponding to some input features, $\bm{x}$, we define matrices $\tilde{\bm{X}}\in\mathbb{R}^{(d+1) \times N}$, $\bm{Y}=\tilde{W}\tilde{\bm{X}}$ and $\bm{Y'}\in\mathbb{R}^{k \times N}$. 
Using these we write the MSE loss as:
\begin{equation}
    L(\params) = \frac{1}{2N} \text{tr} \left( [\bm{Y}-\bm{Y'}][\bm{Y}-\bm{Y'}]^T \right),
\end{equation}
which can be further simplified to
\begin{equation}\label{eq:mse-loss}
    L(\params) = \frac{1}{2} \left( { \text{tr} \left(\tilde{W}A\tilde{W}^T \right) - 2 \text{tr} \left( \tilde{W} B \right) + \text{tr} \left( C \right) } \right),
\end{equation}
where
\begin{align}
    A & = \frac{1}{N}\bm{\tilde{X}}\bm{\tilde{X}}^T, \\
    B & = \frac{1}{N}\bm{\tilde{X}}\bm{Y'}^T, \\
    C & = \frac{1}{N}\bm{Y'}\bm{Y'}^T.
\end{align}

If we express the trajectory parameters $\trajparams$ as a row vector of blocks $\tilde{W}_t$ for each time $t$, $\trajparams\in\mathbb{R}^{k\times\tau(d+1)}$, we can write the partition sum in \er{biased1} as a Gaussian integral
\begin{align}
    \mathcal{Z}_\tau (\sigma,s)=\int D\trajparams \exp [ & -\frac{1}{2} \text{tr}(\trajparams\tilde{A}\trajparams^T) \notag \\
    & + \text{tr} (\trajparams \tilde{B}) - \frac{s}{2}\tau \text{tr}(C) ],
\end{align}
where 
\begin{equation}
    \label{eq:tildeA}
    \tilde{A} = \begin{bmatrix} 
        \frac{\idty}{\sigma^2} + sA & -\frac{\idty}{\sigma^2} & & & 0 \\
        -\frac{\idty}{\sigma^2} & \frac{\idty}{\sigma^2} + 2sA & -\frac{\idty}{\sigma^2} & & \\
        & -\frac{\idty}{\sigma^2} & \ddots  & & \\
        & & & \frac{\idty}{\sigma^2} + 2sA & -\frac{\idty}{\sigma^2} \\
        0 &  &     &   -\frac{\idty}{\sigma^2} & \frac{\idty}{\sigma^2} + sA \\
        \end{bmatrix},
\end{equation}
and 
\begin{equation}
    \label{eq:tildeB}
    \tilde{B} = \begin{bmatrix} 
        sB \\
        \vdots \\
        sB \\
        \end{bmatrix}.
\end{equation}
We see that $\tilde{A} \in \mathbb{R}^{\tau(d+1)\times\tau(d+1)}$ and $\tilde{B} \in \mathbb{R}^{\tau(d+1)\times k}$. Note that when $\tau=1$, $\tilde{A}=sA$, as there is no dependence on $\sigma$. It is easy to integrate over trajectories to obtain the partition sum for the linear perceptron
\begin{align}
    \label{eq:exact-lp-partition-sum}
    \mathcal{Z}_\tau(\sigma,s)= 
        \exp 
        & 
        \left\{ 
            -\frac{1}{2} \text{tr}(\tilde{B}^T\tilde{A}^{-1}\tilde{B}) 
        \right.
    \nonumber \\
    & 
    \;\;\;
    \left.        
    - \frac{1}{2} \log \text{det} \tilde{A}  + \frac{s}{2}\tau \text{tr}(C) 
    \right\} .
\end{align}

\begin{figure*}
    \centering
    \includegraphics[width=\textwidth]{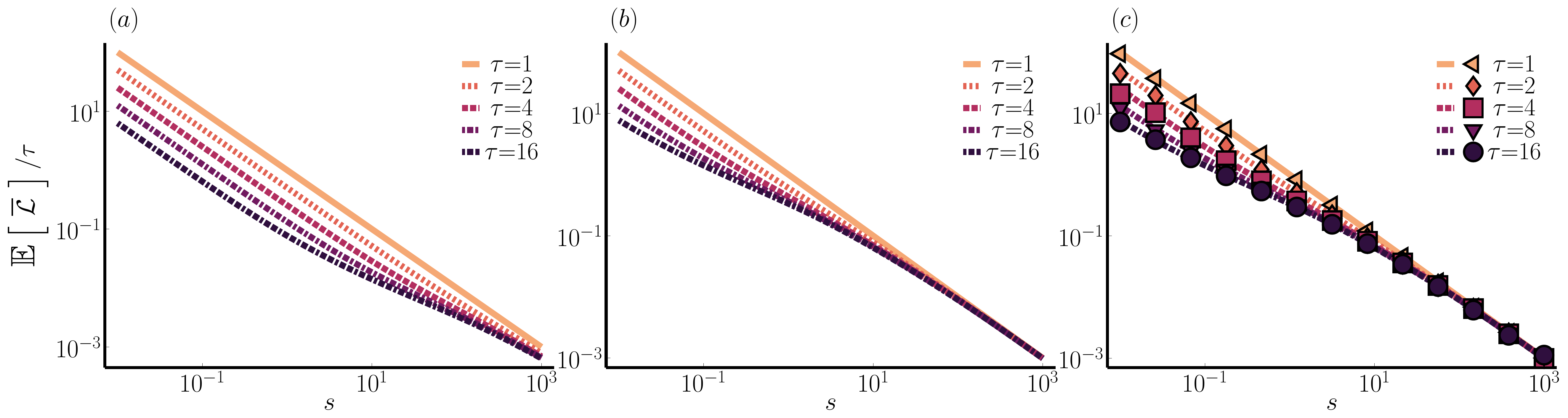}
    \caption{\label{fig:linear-s-ensemble-plots}
        {\bf NNEs as trajectories in the linear perceptron.}
        (a) Exact time-averaged loss, \er{eq:expected-loss-partition-sum}, 
        as a function of $s$ for various trajectory lengths $\tau$ and $\sigma=0.1$. (b) Same for $\sigma=1$. 
        (c) 
        Comparison between exact results (lines) and TPS sampling (symbols). Empirical samples were generated using methods described in Sec.~V, using an adaptive annealing technique in $s$ to decrease convergence time, similar to the process shown in \fr{fig:annealing-scheme}. Full source code to reproduce this data is supplied in \cite{MairTPS2022,MairNNE2022}.
        }
\end{figure*}

The dynamical partition sum \er{Z} is the moment generating function for the time-integrated loss. From this, one can obtain the average time-integrated loss for arbitrary $s$ \cite{Touchette2009,Garrahan2018,Jack2019},
\begin{equation}\label{eq:expected-loss-partition-sum}
    \frac{1}{\tau} \mathbb{E}\left[ \mathcal{L}(\tilde{\omega})\right] = -\frac{1}{\tau} \partial_s \log \mathcal{Z}_\tau (s).
\end{equation}
For the linear perceptron problem, we can compute \er{eq:expected-loss-partition-sum}
directly from \er{eq:exact-lp-partition-sum}. 
The average loss as a function of $s$ is shown in Fig.~\ref{fig:linear-s-ensemble-plots}(a,b), for different values of $\tau$ and two values of $\sigma$. 

We observe the following features: (i) for a given $\tau$, the loss per unit time decreases with increasing $s$; (ii) for fixed $s$, the loss decreases with $\tau$; and (iii) the loss curves are systematically lower in values the smaller $\sigma$. These can be explained from the exact form of the average time-integrated loss of the linear perceptron trajectories. In that expression we find two regimes:
\begin{equation}\label{eq:exact-linear-loss-vs-s}
    \frac{1}{\tau} 
    \mathbb{E}\left[\mathcal{L}(\tilde{\omega})\right] \approx 
    \frac{1}{2}\text{tr}( B^T B ) - \frac{1}{2}\text{tr}({C}) -
    \begin{cases}
        \frac{1}{2s\tau} & s\sigma^2 \ll 1 \\
        \frac{1}{2s} &  s\sigma^2 \gg 1
    \end{cases}
\end{equation}
The first two terms of are constants, given by samples in the dataset. This  specifies the minimum loss achievable, given the data. When $s\sigma^2 \ll 1$, a dependence on $\tau^{-1}$ emerges, which is responsible for the banding in \fr{fig:linear-s-ensemble-plots} for small values of $s$: a higher value of $\tau$ will lead to a lower loss. When $s\sigma^2 \gg 1$, the banding effect becomes negligible, and all values of $\tau$ converge to the same loss per unit time for a given $s$.

The parameter $\sigma$ controls where the banding occurs, see \fr{fig:linear-s-ensemble-plots}: comparing panels (a) and (b) we see that banding persists in (a) much longer than in (b); the $s$ values at which the curves converge, i.e. the loss of advantage of longer trajectories, differs by a factor of $\sigma^2$.
[For $\tau=1$, the conditions in \er{eq:exact-linear-loss-vs-s} become the same, 
and $\sigma$ plays no role; this is the limit of a single NN trained via gradient descent or neuroevolution, as discussed in Sec.~IIIA.]

The interplay between the hyperparameters $s$, $\tau$ and $\sigma$ provides in our trajectory method  
a mechanism for balancing between exploitation versus exploration in the training of the NNE. As we will show below, all these features that we can compute exactly for an ensemble of linear perceptrons generalise qualitatively to more complex architectures: the loss of a NNE obtained from our trajectory approach is reduced by increasing $s$, trajectory length, and decreasing $\sigma$.

\section{Training NNEs with transition path sampling}

In order to sample good NN ensembles from the biased distribution \er{biased1} we use Transition Path Sampling (TPS) \cite{Bolhuis2002}, a form of Monte-Carlo in trajectory space that can be used to converge to sample arbitrary rare trajectories from some unbiased dynamics. It is particularly well suited for sampling dynamics where the biased set is due to tilting by a time-integrated function of the trajectories, see e.g.\ \cite{Hedges2009}. 

TPS proceeds as in standard Monte-Carlo by generalising from configurations to trajectories. In our case we wish to sample from $P_{\sigma,s}(\trajparams)$, \er{biased1}, so the starting point of TPS is to construct transition probabilities $P(\trajparams' | \trajparams)$ which satisfies {\it detailed balance} with respect to that distribution: $P(\trajparams' | \trajparams) P_{\sigma,s}(\trajparams) = P(\trajparams | \trajparams') P_{\sigma,s}(\trajparams')$. This guarantees that the stationary distribution of trajectories generated sequentially using $P(\trajparams' | \trajparams)$ will converge to those sampled from $P_{\sigma,s}(\trajparams)$. The transition probabilities are decomposed into two sub-steps known as the {\it proposal} and {\it acceptance-rejection} steps
\begin{equation}
    \label{eq:transition-decomposition}
    P(\trajparams' | \trajparams)=g(\trajparams' | \trajparams)A(\trajparams', \trajparams),
\end{equation}
where $g(\trajparams' | \trajparams)$ is the conditional probability of proposing a trajectory $\trajparams'$ given a current trajectory of $\trajparams$.
The acceptance probability $A(\trajparams', \trajparams)$ then specifies how likely it is to accept the proposed trajectory, given the current trajectory.
Inserting this decomposition into the detailed balance leads to a relation which the acceptance must satisfy for the combined dynamics to possess the desired detailed balance:
\begin{equation}
\label{eq:acceptance-criteria}
    \frac{A(\trajparams', \trajparams)}{A(\trajparams, \trajparams')}=\frac{P_{\sigma,s}(\trajparams')}{P_{\sigma,s}(\trajparams)}\frac{g(\trajparams | \trajparams')}{g(\trajparams' | \trajparams)}.
\end{equation}
A very common choice for the acceptance ratio which fulfils the above expression is the Metropolis one
\begin{equation}
    \label{eq:acceptance-condition}
    A(\trajparams', \trajparams) = \min \left( 1, \frac{P_{\sigma,s}(\trajparams')}{P_{\sigma,s}(\trajparams)}\frac{g(\trajparams | \trajparams')}{g(\trajparams' | \trajparams)} \right) .
\end{equation}
Given that only ratios of the target distribution appear in the acceptance probability, we can simplify the above expression by choosing proposal moves that satisfy detailed balance with respect to the unbiased probability \eqref{unbiased1}, 
\begin{equation}
    \label{eq:generating-detailed-balance}
    P_{\sigma}(\trajparams')g(\trajparams | \trajparams') = 
	P_{\sigma}(\trajparams)g(\trajparams' | \trajparams).
\end{equation}
Inserting into \eqref{eq:acceptance-condition} we get
\begin{equation}
\label{eq:acceptance-criteria-restricted}
    A(\trajparams', \trajparams) = \min \left( 1, e^{{ -s[\mathcal{L}(\trajparams')-\mathcal{L}(\trajparams)]}} \right) . 
\end{equation}
The above means that trajectories are generated with the unbiased dynamics, 
\er{eq:generating-detailed-balance}, 
and accepted or rejected according to the change in time-integrated loss, \er{eq:acceptance-criteria-restricted}. 

While generating trajectories with the unbiased dynamics is a big simplification, for two arbitrary trajectories the difference in time-integrated loss is extensive in time and (at least) also extensive in number of parameters, making acceptance \eqref{eq:acceptance-criteria-restricted}, in general, exponentially small. 
For that matter, most of the art in TPS is to design moves that both satisfy detailed balance in trajectory space and make acceptance efficient.

\subsection{Generating dynamics: shooting plus Brownian bridges}

A very common choice for generating trajectory moves in TPS is {\em shooting} \cite{Bolhuis2002}, which involves choosing a fixed time in the trajectory and a direction, forwards or backwards, and evolving with the original unbiased dynamics until reaching the end of the trajectory (with detailed balance, a backward shooting move can be generated forwards and time-reversed). 
Shooting with the original dynamics \er{unbiased1} is outlined in Algorithm \algr{alg:shooting}. 
Under shooting, $g$ obeys \er{eq:generating-detailed-balance} and 
therefore \er{eq:acceptance-criteria-restricted} is the corresponding acceptance probability. As
shooting leaves a portion of the trajectory unchanged, one need only calculate the change in loss of the modified portions of the trajectory.
The shooting method is sketched in 
\fr{fig:perturbation_examples}(a,b).

\begin{algorithm}[H]
	\caption{\label{alg:shooting} Shooting TPS}
	\begin{algorithmic}[1]
		\State \textbf{input} Current trajectory $\trajparams$
		\State \textbf{parameters} Variance of the Gaussian noise $\sigma^2$
		\State Choose $t$ uniformly from $[1, \tau]$, where $t$ is an integer
		\State Choose a direction randomly $\kappa \in \{-1, 1\}$
		\State Initialise $\trajparams' \gets \trajparams$
		\While {$t+\kappa \geq 1$ \textbf{and} $t+\kappa \leq \tau$}
		    \State Sample all elements of $\Delta\params'$ i.i.d.\ from $\mathcal{N}(0, \sigma^2)$ 
		    \State ${\params'}_{t+\kappa} \gets {\params'}_{t} + \Delta \params'$
		    \State $t \gets t + \kappa$
		\EndWhile
		\State \textbf{output} $\trajparams'$
	\end{algorithmic}
\end{algorithm}

\begin{figure*}
	\centering
	\includegraphics[width=\textwidth]{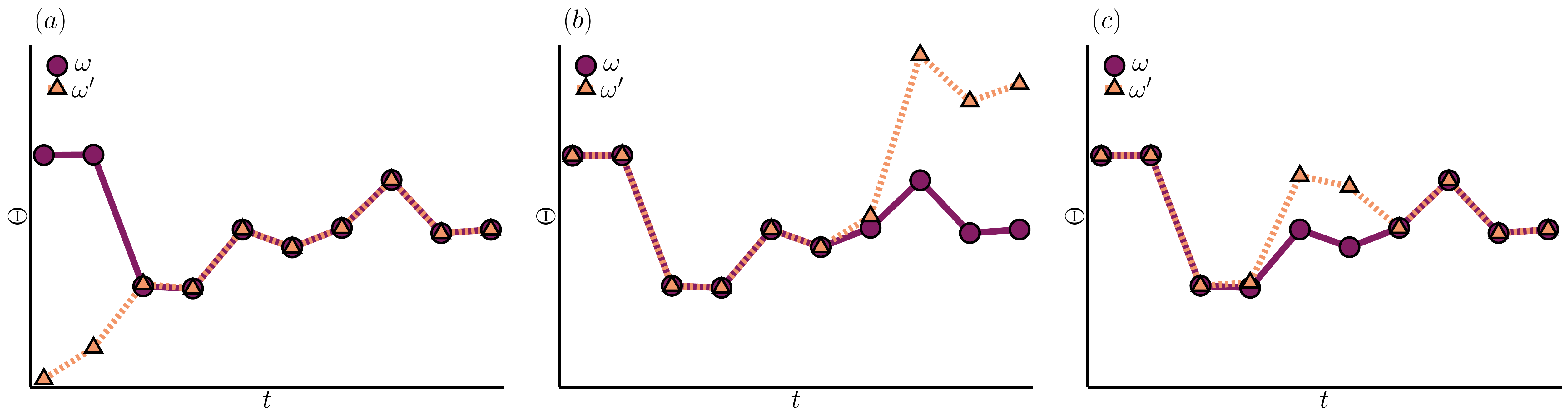}
	\caption{\label{fig:perturbation_examples} 
	{\bf TPS scheme.}
	The original trajectory is  $\omega$ and the proposed new trajectory is $\omega'$. (a) Backwards shooting, fixing $t\geq4$.
	(b) Forwards shooting, fixing $t\leq 6$.
	(c) Brownian bridge, fixing $t \leq 2$ and $t \geq 7$.}
\end{figure*}

A common issue with shooting is that it is difficult to generate successful updates towards the centre of a trajectory. 
The reason is that when shooting from the bulk of the trajectory, the difference in time-integrated loss between the proposed and current trajectories scales with the trajectory length, making acceptance exponentially suppressed with time. 
To address this issue, we exploit the fact that the unbiased dynamics we use to generate moves is Brownian: we supplement  shooting with moves generated by {\em Brownian bridges}, see e.g.\ \cite{Grela2021,De-Bruyne2021}, that is, 
a proposed move consists of replacing a portion of the trajectory by a bridge between the same initial and final points of the replaced segment, see sketch in
\fr{fig:perturbation_examples}(c). 
By controlling the time extent of the bridge, we can attenuate the loss difference in \er{eq:acceptance-criteria-restricted} thus enhancing acceptance. Details 
on the Brownian bridges is given in  the Appendix.
The key formulae are as follows: if the portion of the trajectory to replace is between times $t_1$ and $t_2$, keeping $x_{t_1}$ and $x_{t_2}$ fixed, by generating a bridge with the dynamics
\begin{align}
	P_{\rm B}\left(x_t|x_{t-1},t\right)=\frac{e^{-\frac{\left[x_t-\mu(x_{t-1},t)\right]^2}{2v(t)}}}{\sqrt{2\pi v(t)}},
\end{align}
with {\em time-dependent} mean and variance
\begin{align}
	\label{eq:bridge-mean}
	\mu(x,t)=\frac{x_{t_2}+(t_2-t)x}{t_2-t+1},\\
	\label{eq:bridge-variance}
	v(t)=\sigma^2 \frac{t_2-t}{t_2-t+1}.
\end{align}

We demonstrate how a bridge is generated in \algr{alg:bridges}. One has the choice of how to generate $t_1$ and $t_2$. In our simulations, we favour choosing $t_1$ uniformly from $[1, \tau-2]$ and then setting $t_2=t_1+2$, to give a bridge of a single time step, so there is only one updated state.

\begin{algorithm}[H]
	\caption{\label{alg:bridges} Brownian bridges for TPS}
	\begin{algorithmic}[1]
		\State \textbf{input} Current trajectory $\trajparams$
		\State \textbf{parameters} Variance of the Gaussian noise $\sigma^2$
		\State Choose $t_1$ and $t_2$ uniformly from $[1, \tau]$, where $t_1,t_2 \in \mathcal{Z}$ and $t_2>t_1$.
		\State Initialise $\trajparams' \gets \trajparams$
		\State Initialise $t \gets t_1+1$
		\While {$t < t_2$}
            \State Calculate mean $\bm{\mu}_t$ for $(\params'_{t-1}, t)$ from \er{eq:bridge-mean} 
			\State Calculate variance $v(t)$ from \er{eq:bridge-variance}
		    \State Sample $\theta'_a$ from $\mathcal{N}[(\bm{\mu}_t)_a, v_t]$ for all components $a$ 
		    \State ${\params'}_{t} \gets {\params'}$
		    \State $t \gets t + 1$
		\EndWhile
		\State \textbf{output} $\trajparams'$
	\end{algorithmic}
\end{algorithm}

One cannot guarantee ergodicity in trajectory space using only bridges, as it requires two ends to be fixed, meaning the end of the trajectories will not be modified. Instead, we use a combined approach consisting of choosing the shooting algorithm (\algr{alg:shooting}) with probability $p_{\text{shoot}}$ or the bridge algorithm (\algr{alg:bridges}) with $1-p_{\text{shoot}}$.
For shorter trajectories $\tau\leq 4$, we choose $p_{\text{shoot}}=1$, as the trajectories are not long enough for centre trajectory updates to become inefficient. When $\tau > 4$, we set $p_{\text{shoot}}=\frac{2}{\tau}$ and modify the shooting algorithm to only shoot forwards from $t=\tau-1$ or backwards from $t=2$ with equal probability, and choose bridges that only alter a single time in the trajectory as described earlier. This choice was to improve acceptance, and ensure that each model in the trajectory had an equal probability of being mutated.

\section{Numerical Results}

\subsection{Linear perceptron}

As an elementary demonstration of our TPS scheme, we applied it to the linear perceptron model of Sec.~\ref{sec:exact-linear-perceptron}. Figure \ref{fig:linear-s-ensemble-plots}(c)
shows that TPS reproduces the exact results for the average time-integrated loss as a function of $s$.

The quantity of interest is the expected time-averaged loss of the NNE. We can estimate this quantity as a running average of the TPS iterations, or ``epochs'' of the training,
\begin{equation}
    \label{eq:sample-time-avg-loss}
    \mathbb{E}\left [ \mathcal{L}(\trajparams) \right ] \approx 
    \frac{1}{M}
    \sum_{m={M}_{\text{rel}}}^{{M}_{\text{rel}}+{M}} 
    \mathcal{L}(\trajparams^{(m)}),
\end{equation}
where $\trajparams^{(m)}$ is the parameter trajectory at TPS epoch $m$. In the above, the time-integrated loss is an empirical average over ${M}$ epochs, 
calculated after allowing TPS to relax for ${M}_{\text{rel}}$ epochs, large enough so that TPS converges to stationarity. In the limit of ${M}\rightarrow \infty$ \er{eq:sample-time-avg-loss} becomes an equality. As is standard practice, we check for TPS relaxation empirically. A common technique for speeding up convergence is to anneal the $s$ parameter, starting at a low $s$ (i.e.\ close to the unbiased dynamics), and progressively increasing $s$ towards the desired value; see \fr{fig:annealing-scheme}.

\begin{figure}
	\centering
	\includegraphics[width=0.5\textwidth]{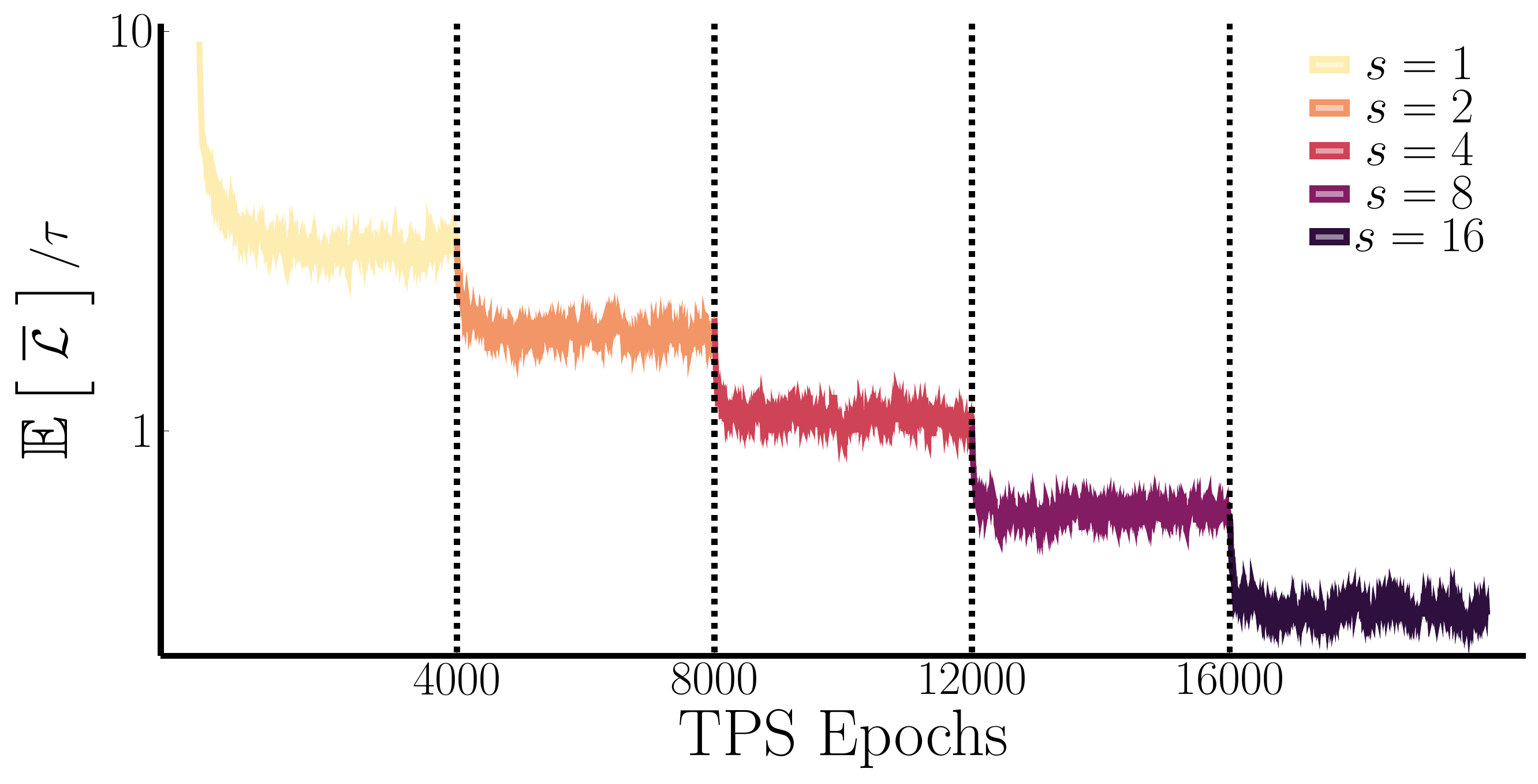}
	\caption{\label{fig:annealing-scheme} 
    {\bf TPS annealing.}
    Training of the NNE for the linear perceptron through stepwise decrease of $s$,     
    for $\tau=8$ and $\sigma=1.0$.}  
\end{figure}

\subsection{Two-dimensional loss landscape}

To illustrate how TPS works on a trajectory representing a model, and in particular how biasing the diffusive dynamics allows for sampling low loss trajectories, we consider a simple, yet non-trivial, toy problem where the model has only two parameters. 

We consider the Hummelblau's function:
\begin{equation}
	h(x, y) = (x^2 + y - 11)^2 + (x + y^2)^2, 
\end{equation}
and choose $x$ and $y$ to be within $[-5, 5]$. The aim is to train an ensemble of models, where each model has parameters $\theta=(x, y)$ and $h(x,y)$ is the loss.  We can then construct a trajectory, $\trajparams$, which is initialised under Gaussian dynamics with $\sigma=1.0$, with $\theta_0=(x_0, y_0)$ randomly chosen from $[-1, 1] \times [-1, 1] $. As before, we choose the observable of a trajectory of length $\tau$ to be the time-integrated loss, ${\mathcal{L}}(\trajparams) = \sum_{t=1}^\tau h(x_t, y_t)$. 

\begin{figure}
    \centering
    \includegraphics[width=0.5\textwidth]{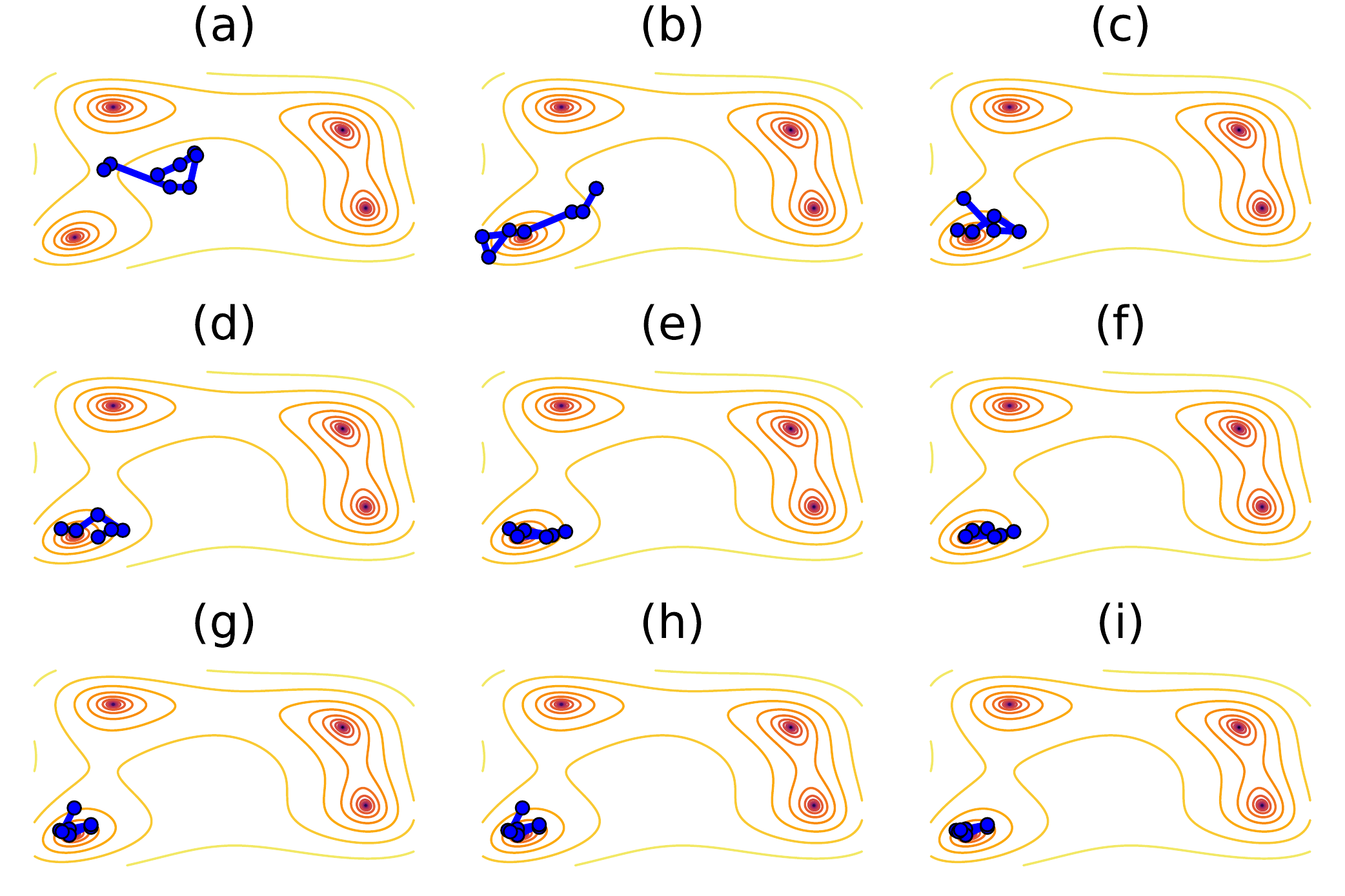}
    \caption{\label{fig:hummelblau-iterations} 
    {\bf Trajectory evolution under TPS.}
    Progression of the NNE training 
    for the 2D loss landscape. Snapshots of the trajectory every $50$ TPS epochs.}
\end{figure}

Figure \ref{fig:hummelblau-iterations} shows the evolution of the ensemble trajectory under TPS. The first snapshot in panel (a) is the randomly initialised trajectory, and progress is shown every $50$ TPS iterations. We see that as TPS progresses, the trajectory evolves towards being localised in the minimum of the loss landscape. This represents the training of the ensemble by means of trajectory sampling.

In this simple example, it is easy to visualize two features of the training. First, loss landscapes can have a natural length scale for each dimension, which justifies a choice of $\sigma$ to enable a high enough acceptance rate. Second, both $s$ and $\sigma$ dictate the evolution of a trajectory under TPS, and therefore the training of the ensemble.
A lower value of $s$ and a higher value of $\sigma$ means that the trajectory is free to spread out across the loss landscape, fairly unimpeded. A higher value of $s$ and a lower value of $\sigma$ restricts a trajectory to cover a small range of the loss landscape, making the trajectory much more likely to be localised in a local minimum. 

\subsection{MNIST classification}

As a final example, we consider a standard ML classification task. 
In this case the relevant loss function is the mean cross entropy loss over the data. For a single NN this reads
\begin{equation}
    \label{eq:cross-entropy-loss}
    L(\params) = -\frac{1}{N_\text{sp}}\sum_{i=1}^{N_\text{sp}} \sum_{j=1}^{N_\text{cl}} {\delta_{z_i, j} \ln y_j(x_i; \params)},
\end{equation}
where $x_i$ and $z_i$ represent the features and class label index of the $i^\text{th}$ data point (in a dataset of size $N_\text{sp}$), 
and $y_j(x_i; \params)$ is the probability of selecting class $j$ 
(out of $N_\text{cl}$ classes)
using the NN parameterised by $\params$. This probability is usually calculated using a softmax layer on the end of the NN. If a model predicts that all classes have an equal probability, then the mean cross entropy loss is $\ln({{N}_{\text{cl}}})$.

We now consider the classification of handwritten digits from the 
MNIST \cite{deng2012mnist} 
dataset, a standard benchmark for many ML algorithms.
Examples of each of the $10$ digits in MNIST are shown in \fr{fig:mnist-all-graphs}(a). MNIST classification is a typical example of a problem where one would apply the cross entropy loss function 
\eqref{eq:cross-entropy-loss}
to train a NN.
Being high dimensional, this is also a good problem to test 
our method of producing trained NNEs via trajectory sampling. 

As we are interested in studying the training process under our ensemble method, we have simplified the problem by restricting the dataset to $2048$ samples, down from the usual $60,000$. The distribution of each digit is uniform in the sampled training set. The basic architecture of each NN in the ensemble uses a mix of convolution and fully connected layers, based on LeNet\cite{Lecum1998}. The aim is to train the NNE to output a probability of selecting a digit, based on the softmax distribution of the final output layer of the NN. The cross-entropy loss function, \er{eq:cross-entropy-loss}, is calculated using the logits (the input to the softmax activation) for numerical stability. 

In a classification problem, one uses the loss function as a proxy for accuracy. These two measures are highly correlated as a low loss likely correlates with high accuracy. Effectively, to achieve a high accuracy on the training dataset, there is a low loss region which should be reached. For training the NNE via trajectory sampling, we tuned the range of $s$ such that the models converged to a low enough loss region, which corresponded to the full range of accuracies. For the MNIST problem, this range was between $s=5$ and $s=50$. To provide banding in the chosen $s$ domain, we used $\sigma=0.05$ to effectively shift the banding region into these higher values of $s$. Reducing the value of $\sigma$ provides the additional benefit of making smaller updates in TPS, leading to a higher acceptance rate and speeding TPS convergence.

\begin{figure*}
    \centering
    \includegraphics[width=\textwidth]{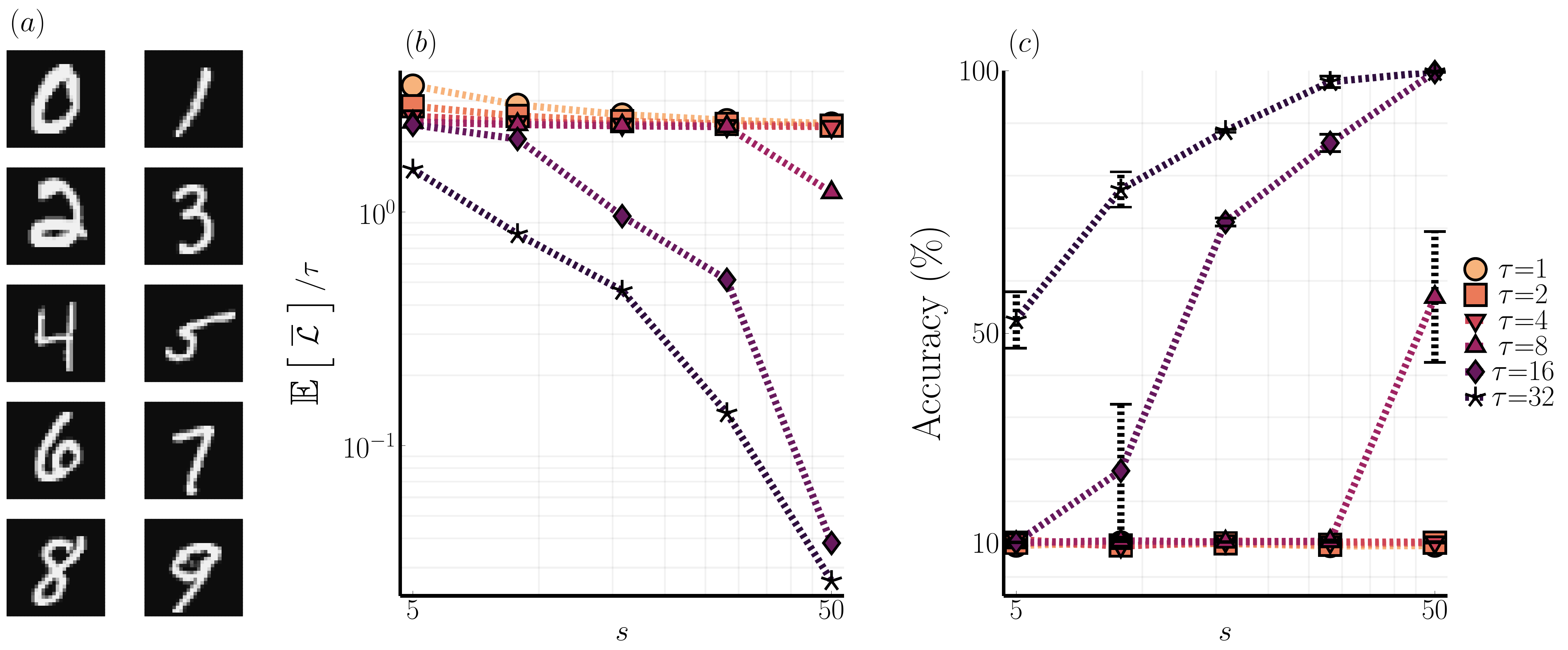}
    \caption{\label{fig:mnist-all-graphs} 
    {\bf NN ensembles training for MNIST.}    
    (a) Representative images for the $N_{\text{cl}} = 10$ classes of  the MNIST dataset. Each digit is a $28\times 28$ greyscale image, with the brightness of each pixel encoded between $0.0$ (black) and $1.0$ (white). (b) Time-averaged cross entropy loss for different values of $s$ and $\tau$, for $\sigma=0.05$. (c) Time-averaged mean accuracy from the NNE at the final TPS epoch of the training, 
    for the same hyperparameters of (b). The accuracy expected for a random NN is around $10\%$, as there are $10$ classes in the MNIST problem. This corresponds to the loss plateau at low $s$ for small $\tau$.}
\end{figure*}

We ran a combination of shooting and bridging TPS, as detailed previously, for a number of independent trajectories until the time-averaged loss function appeared numerically converged, similar to what can be seen in \fr{fig:annealing-scheme}. Figure \ref{fig:mnist-all-graphs}(b) shows the time-integrated loss per unit time as a function of $s$ of the  trained NNEs. We can see an analogous banding to that of the simple linear perceptron: larger values of $s$ have lower ensemble loss, and this is significantly enhanced by increasing the trajectory length. 

In \fr{fig:mnist-all-graphs}(b) we also see differences with the perceptron arising from the non-polynomial form of the loss, \er{eq:cross-entropy-loss}. 
First is the effect of the minimum possible value of the loss that the chosen NN architecture can reach. This minimum should be approached by curves as $s \rightarrow \infty$. In 
\fr{fig:mnist-all-graphs}(c) we show that this tends towards $100\%$ accuracy of the NNE, already achieved for $s=50$ and $\tau=32$. The second difference is in the small $s$ regime, where the loss function exerts less influence over the sampling. 
Since under the unbiased dynamics 
each output logit is randomly distributed in the long time limit, for vanishing $s$ 
the mean cross-entropy loss per unit time will converge to $\ln({{N}_{\text{cl}}})$, provided that the classes are balanced. This produces an upper loss plateau in the dynamics. Increasing the loss beyond this point requires a model to be ``intelligently wrong'', by lowering the probability of choosing the correct answer beyond uncorrelated random chance. 
In contrast, for a polynomial loss function as in the linear perceptron case, the loss diverges as $s \to 0$, something that does not occur in a classification problem like this MNIST. 

The highest value of $s$ we used for the MNIST problem is large enough to train the longer trajectories to very low losses, which correspond to very high accuracies, see \fr{fig:mnist-all-graphs}(c). While we could in principle reduce the losses to arbitrarily small values, this often causes overfitting, reducing the ability of the models to generalise to unseen examples, see e.g.~\cite{Goodfellow2016,Mehta2019}. Additionally, there is significant cost in using higher values of $s$, as they reduce the acceptance rate, which can significantly increase training time and cost. We demonstrated that our method is capable of training a standard convolutional neural network to high train accuracy on a subset of the MNIST problem, as seen in \fr{fig:mnist-all-graphs}(c).
\section{Conclusions}

Here, we have presented a method to train neural network ensembles using trajectory sampling techniques more often applied in the statistical mechanics of non-equilibrium systems. In our approach the set of neural networks that form the NNE corresponds to the sequence of configurations of the NN parameters that are visited in a stochastic trajectory. By biasing trajectories to have low time-integrated loss we showed we could train NNEs to perform well in standard machine learning tasks such as MNIST classification. For concreteness, we focused on trajectories from dynamics which is discrete in time, continuous in space, and where the changes at each time step are synchronous, in the sense that all parameters can be updated simultaneously. None of these is a requirement: the underlying dynamics can equally be taken to be continuous in time, as in a continuous-time Markov chain or in diffusions, and parameter updates do not need to be Gaussian; in such a case, the rest of the approach for training and sampling would be essentially the same as the one above.  

Our trajectory NNE method has to be compared to those based on gradient descent, which focus on the shortest route to a set of parameters for a NN which locally minimise the loss. In contrast, our method is in the spirit of thermal sampling, where low loss configurations are searched by controlling a parameter (such as temperature in thermal annealing, or $s$ in our trajectory method) which being coupled to the quantity of interest (energy or time-integrated loss) pushes towards low values, balanced with exploring state or trajectory space. 
An obvious drawback of gradient descent is its inability to escape local traps, and this is the reason that modern ML supplements it with noise and inertia to make it efficient. Gradient-free methods like ours are less sensitive to local trapping, which is especially prominent in smaller and non-over-parameterised NNs, with their physical interplay between minimising the observable and maximising the entropy playing an analogous role as the exploit/explore trade-off of ML learning techniques.

The above, together with the ability of NNEs to reduce overfitting, suggest to us that the approach proposed here will be most useful when constructing ensembles of smaller models as compared to a single, larger, NN. 
While in this paper we focused only on introducing the trajectory NNE method and showing its viability, we hope to report in future work on systematic comparison on performance and training cost between our NNEs and a single NN.  

Our approach relies on converging to the stationary state in trajectory space determined by the hyperparameters $s$, $\sigma$ and $\tau$. Their meaning is clear: $s$ controls
the level of the ensemble loss, with larger $s$ leading to lower overall loss; $\sigma$ controls exploration, with larger $\sigma$ allowing for larger fluctuations in the trajectory of models; and $\tau$ determines the size of the ensemble. While increasing $s$, decreasing $\sigma$, and increasing $\tau$ all reduce the NNE loss, the ability to control the three hyperparameters separately provides much flexibility for the training. The same applies to numerical ``convergence'': while ideally one would like to sample trajectories from the stationary state of \er{biased1}, in practice all that is required is that the TPS iterations reach trajectories of low enough loss for the problem at hand. 

The effectiveness of TPS relies on a reasonable acceptance for proposed trajectory updates. In general, acceptance is exponentially suppressed in trajectory length and size of the system. We resolved the exponential in time problem by proposing bridge moves which are localised in time. For a NN system where the loss is a fully connected function of the weights, the exponential in size cost problem is more difficult to solve. For training NNEs with larger NN constituents this might become a limiting factor. A related issue is that of batching the data when calculating the loss: in the ML/statistical mechanics analogy, under learning dynamics the parameters of the NN are the fluctuating variables, while the training dataset is like quenched disorder in the interactions defining the loss. Using data batches to calculate the loss (a standard trick in ML that gives rise to stochastic gradients), is equivalent to having (slow) fluctuating disorder, something which has not been studied in as much detail in the context of trajectory sampling. Further integration of ML and non-equilibrium ideas will help improve the trajectory NN ensemble even further. We hope to report on such developments in the future.

\begin{acknowledgments}
    We acknowledge financial support from the Leverhulme Trust Grant RPG-2018-181 and University of Nottingham grant no.\ FiF1/3. DCR was supported by funding from the European Research Council (ERC) under the European Union’s Horizon 2020 research and innovation programme (Grant agreement No. 853368). Simulations were performed using the University of Nottingham Augusta HPC cluster, and the Sulis Tier 2 HPC platform hosted by the Scientific Computing Research Technology Platform at the University of Warwick. Sulis is funded by EPSRC Grant EP/T022108/1 and the HPC Midlands+ consortium. We thank the creators and community of the Julia programming language \cite{julialang2017}, and acknowledge use of the packages \texttt{CUDA.jl} \cite{besard2018juliagpu,besard2019prototyping}, \texttt{Flux.jl} \cite{Flux.jl-2018, innes:2018}, \texttt{Plots.jl} \cite{plotsjl2022} and \texttt{ForwardDiff.jl} \cite{RevelsLubinPapamarkou2016}. 
    Our TPS implementation package is available through GitHub, \texttt{TransitionPathSampling.jl} \cite{MairTPS2022}, together with the source code to generate the figures and results in the paper \cite{MairNNE2022}.
\end{acknowledgments}

\appendix

\appendix

\section*{Appendix: Exact bridge dynamics for a discrete time gaussian process}
\label{app:bridge_dynamics}

Consider an original dynamics of a position $x\in\mathbb{R}$ given by gaussian movements with variance $v$ at each time step
\begin{align}
	P(x'|x)&=\int\frac{dw}{\sqrt{2v\pi}}e^{-\frac{w^2}{2v}}\delta(x'-x-w),\\
	&=\frac{e^{-\frac{(x'-x)^2}{2v}}}{\sqrt{2v\pi}},
\end{align}
and initial probability distribution $P(x_0)=\delta(x_0-x_i)$.
The probability of an individual trajectory $\omega_0^T=\{x_t\}_{t=0}^T$ of length $T$ is given by
\begin{align}
	P\left(\omega_0^T\right)=\prod_{t=1}^TP(x_t|x_{t-1})P(x_0).
\end{align}
We seek a dynamics which produced, with the correct relative probabilities, the subset of trajectories given by this dynamics such that they all end at $x_T=x_f$, so-called bridge trajectories.
That is, we seek a Markovian dynamics which generates trajectories with probability
\begin{align}
	P_B\left(\omega_0^T\right)=\frac{\delta(x_T-x_f)P\left(\omega_0^T\right)}{\sum_{\omega_0^T}\delta(x_T-x_f)P\left(\omega_0^T\right)}.
\end{align}
We can expand this trajectory probability using the probabilistic chain rule as
\begin{align}
	P_B\left(\omega_0^T\right)=\prod_{t=1}^TP_B(x_t|\omega_0^{t-1})P_B(x_0),
\end{align}
solving for these probabilities, which will turn out to be Markovian, iteratively.
For the last time step we find
\begin{align}
	P_B\left(x_T|\omega_0^{T-1}\right)
	&=\frac{P_B\left(\omega_0^{T}\right)}{P_B\left(\omega_0^{T-1}\right)}\\
	&=\frac{\delta(x_T-x_f)P\left(\omega_0^T\right)}{\int dx_T \delta(x_T-x_f)P\left(\omega_0^T\right)}\\
	&=\frac{\delta(x_T-x_f)P\left(x_T|x_{T-1}\right)}{\int dx_T \delta(x_T-x_f)P\left(x_T|x_{T-1}\right)}\\
	&=\delta(x_T-x_f):=P_B\left(x_T|x_{T-1},T\right),
\end{align}
while for the rest we find
\begin{align}
	P_B\left(x_t|\omega_0^{t-1}\right)
	&=\frac{P_B\left(\omega_0^{t}\right)}{P_B\left(\omega_0^{t-1}\right)}\\
	&=\frac{\sum_{\omega_{t+1}^T}\delta(x_T-x_f)P\left(\omega_0^T\right)}{\sum_{\omega_t^T} \delta(x_T-x_f)P\left(\omega_0^T\right)}\\
	&=\frac{g(x_t,t)P(x_t|x_{t-1})}{g(x_{t-1},t-1)}\label{eq:gauge-transformed}\\
	&:=P_B\left(x_t|x_{t-1},t\right)
\end{align}
where we have defined
\begin{align}
	g(x_t,t)=\sum_{\omega_{t+1}^T}\delta(x_T-x_f)P\left(\omega_{t+1}^T|x_t\right).
\end{align}
Finding these scaling factors thus returns the desired Markovian dynamics.
First, for $g(x,T-1)$ we find
\begin{align}
	g(x,T-1)=\int dx'\delta(x_T-x_f)P(x'|x)=\frac{e^{\frac{(x_f-x)^2}{2v}}}{\sqrt{2v\pi}}.
\end{align}
To find the rest, we note that these scaling factors satisfy an inductive equation, a non-linear Bellman equation, due to the normalization of the dynamics
\begin{align}
	g(x,t-1)=\sum_{x'}P(x'|x)g(x',t),\label{eq:bellman}
\end{align}
which we can solve inductively.
We consider the ansatz
\begin{align}
	g(x,T-i)=\frac{e^{\frac{1}{v}\left(a_ix_f^2+b_ix^2+c_ix_fx\right)}}{n_i\sqrt{2v\pi}},
\end{align}
for $i\geq1$, with $n_1=1$, $a_1=-\frac{1}{2}$, $b_1=-\frac{1}{2}$, $c_1=1$.
Using the Bellman equation \eqref{eq:bellman} we thus find
\begin{align}
	n_{i+1}=n_i\sqrt{1-2b_i},\\
	a_{i+1}=a_i-\frac{c_i^2}{4b_i-2},\\
	b_{i+1}=-\frac{1}{2}-\frac{1}{4b_i-2},\\
	c_{i+1}=-\frac{2c_i}{4b_i-2},
\end{align}
which are solved by
\begin{align}
	n_{i}=\sqrt{i},\\
	a_{i}=-\frac{1}{2i},\\
	b_{i}=-\frac{1}{2i},\\
	c_{i}=\frac{1}{i}.
\end{align}
Substituting into \eqref{eq:gauge-transformed} and rearranging we find a Gaussian with time and position dependent mean and time dependent variance
\begin{align}
	P_B\left(x_t|x_{t-1},t\right)=\frac{e^{-\frac{\left[x_t-\mu(x_{t-1},t)\right]^2}{2v(t)}}}{\sqrt{2\pi v(t)}},
\end{align}
where
\begin{align}
	\mu(x,t)=\frac{x_f+(T-t)x}{T-t+1},\\
	v(t)=v\frac{T-t}{T-t+1}
\end{align}

\bibliography{references}

\end{document}